\documentclass[12pt]{article}
\usepackage{amsbsy,graphicx}

\def\al{\alpha}
\def\be{\beta}
\def\ga{\gamma}
\def\de{\delta}
\def\ep{\epsilon}

\def\la{\lambda}
\def\th{\theta}

\def\vp{\varphi}

\def\Ga{\Gamma}

\def\pd{\partial}
\def\del{{\bf\nabla}}

\newcommand{\beq}{\begin{equation}}
\newcommand{\eeq}{\end{equation}}
\newcommand{\bea}{\begin{eqnarray*}}
\newcommand{\eea}{\end{eqnarray*}}
\newcommand{\beaq}{\begin{eqnarray}}
\newcommand{\eeaq}{\end{eqnarray}}
\newcommand{\vph}{\boldsymbol{\varphi}}
\newcommand{\vrh}{\boldsymbol{\rho}}
\newcommand{\val}{\boldsymbol{\alpha}}

\newcommand{\va}{{\bf a}}
\newcommand{\vP}{{\bf P}}

\newcommand{\bfe}{{\bf e}}

\begin{document}
\begin{flushright}SNUTP01-037\\hep-th/0110218\end{flushright}
%\begin{flushright}SNUTP01-037\\KIAS-P99055\end{flushright}
\vspace{10mm}
\centerline{\Large \bf Reflection Amplitudes of Boundary Toda Theories}
\centerline{\Large \bf and Thermodynamic Bethe Ansatz}
\vskip 1cm
\centerline{\large Changrim Ahn$^{1}$, 
Chanju Kim$^{2}$ and Chaiho Rim$^{3}$}
\vskip 1cm
\centerline{\it$^{1}$Department of Physics, Ewha Womans University}
\centerline{\it Seoul 120-750, Korea}
\vskip .5cm
\centerline{\it $^{2}$ Department of Physics and Center for Theoretical
Physics, Seoul National University}
\centerline{\it Seoul 151-747, Korea}
\vskip .5cm
\centerline{\it $^{3}$ Department of Physics, Chonbuk National University}
\centerline{\it Chonju 561-756, Korea}
\vskip 1cm
\centerline{\small PACS: 11.25.Hf, 11.55.Ds}
\vskip 2cm
\centerline{\bf Abstract}
We study the ultraviolet asymptotics in $A_n$ affine Toda theories
with integrable boundary actions. 
The reflection amplitudes of non-affine Toda theories
in the presence of conformal boundary actions have been
obtained from the quantum mechanical reflections of the wave functional 
in the Weyl chamber and used for the quantization conditions 
and ground-state energies.
We compare these results with the thermodynamic Bethe ansatz derived from
both the bulk and (conjectured) boundary scattering amplitudes.
The two independent approaches match very well and provide the 
non-perturbative checks of the boundary scattering amplitudes for 
Neumann and $(+)$ boundary conditions.
Our results also confirm the conjectured boundary vacuum energies
and the duality conjecture between the two boundary conditions.

\newpage
\section{Introduction}
%%%%%%%%%%%%%%%%%%%%%%%%%%%%%%%%%%%%%%%%%%%%%%%%%%%%%%%%
A large class of massive 2D integrable quantum field theories (IQFTs)
can be considered as perturbed conformal field theories (CFTs) 
\cite{sasha}. 
The ultraviolet (UV) behavior of these  IQFTs is encoded in the
CFT data while their long distance properties are defined by the S-matrix
data. 
If the basic CFT admits the representation of the primary fields
of full symmetry algebra in terms of the exponential fields,
the CFT data include ``reflection amplitudes". 
These functions define the
linear transformations between different exponential fields, corresponding
to the same primary field. Reflection amplitudes  play a crucial role
for the calculation of the one-point functions \cite{FLZZ} as well as for 
the description of the zero-mode dynamics \cite{ZamZam, AKR, AFKRY} in  
integrable perturbed CFTs. 
In particular, the zero-mode dynamics determines the UV 
asymptotics of the ground state energy $E(R)$ (or effective central charge
$c_{\rm eff}(R)$) for the system on the circle of size $R$. The function
$c_{\rm eff}(R)$ admits in this case the UV series expansion in the inverse
powers of $\log(1/R)$. The solution of the quantization condition for the
vacuum wave function (which can be written in terms of the reflection 
amplitudes), supplemented with the exact relations between the parameters
of the action and the masses of the particles determines 
all logarithmic terms in this UV expansion.

The effective central charge $c_{\rm eff}(R)$ in IQFT can be also calculated 
independently from the S-matrix data using the TBA method \cite{TBA}.
At small $R$ its asymptotics can be compared with that following from the CFT
data. In the case when the basic CFT is known the agreement of both approaches
can be considered as nontrivial test for the S-matrix amplitudes in IQFT.
The corresponding analysis based on the both approaches was previously done
for the sinh-Gordon \cite{ZamZam}, supersymmetric sinh-Gordon,
Bullough-Dodd \cite{AKR} models, simply-laced affine Toda
field theories (ATFTs) \cite{AFKRY} and nonsimply-laced ATFTs
\cite{ABFKR}.

In this paper we extend this method to the ATFTs with integrable
boundary actions. 
IQFTs with the integrable boundary actions can also be interpreted as 
boundary CFTs perturbed by both bulk and boundary
operators \cite{GhoZam}.
The boundary ATFTs are the non-affine Toda theories (NATTs) with
boundary perturbed by both bulk and boundary operators associated with
the affine roots.
These models become increasingly interesting due to their potential 
applicability to condensed matter systems.  
For IQFTs with boundary, a new physical quantity called 
``boundary $S$-matrix''
\footnote{This object is also called ``boundary reflection amplitude'' 
in some literature. 
Instead we use this terminology to avoid confusion with the boundary version
of reflection amplitude which will be introduced later.}
satisfies the boundary Yang-Baxter equations and associated
bootstrap equations.
These equations determine the boundary $S$-matrices upto CDD-like 
factors and most cases are without direct relations with the boundary actions.
It is an important issue to relate these two informations.
Differently from the bulk, even perturbative checks for 
the boundary $S$-matrices are very complicated because of
the half-line geometry.
One of our main results in this paper is to provide such
non-perturbative confirmation of the proposed boundary $S$-matrices of 
the $A_n$ ATFTs.

For this purpose, we work out the boundary version of the TBA
equations for these models to obtain the ground state energy.
The effects of the boundary $S$-matrices are encoded into the fugacity 
of the TBA equations.
To describe the zero-mode dynamics, we obtain the
``boundary reflection amplitudes'' of the NATTs by considering 
reflections of the wave functional inside the Weyl chamber.
Our results are exactly the same as those obtained by functional
relations in \cite{Fateev}.
The quantization conditions and effective central charges of the ATFTs
can be obtained from these reflection amplitudes. 
We show that these two independent results match upto high accuracy
for $A_n$ ATFTs and corresponding boundary actions.
While our analysis is valid in the UV region, it is noticed that two
results agree well even upto $R \sim {\cal O}(1)$ if exact vacuum
energies are considered. 
There are two contributions to the vacuum energies, one from the 
bulk term which is proportional to $R^2$ and the other from the boundary
term proportional to $R$.
Both contributions become significant as $R$ increases.
Accurate agreement upto this scale provides nonperturbative
check of the boundary vacuum energies conjectured in \cite{Fateev}. 

The rest of the paper is organized as follows. In sect.2 we introduce
the simply-laced ATFTs with integrable boundary actions along with 
the boundary TBA equations based on the bulk and boundary $S$-matrices
corresponding to the Neumann and $(+)$ boundary conditions (see below).
In sect.3 we consider $A_1$ NATT, namely the 
Liouville field theory (LFT) with boundary, and estabilish the wave 
functional description in terms of the boundary reflection amplitude of 
the LFT with boundary.
Using this result, we derive the boundary reflection amplitudes for
the simply-laced NATTs.
In sect.4 we analyze the quantization conditions and UV asymptoics of
ground state energies and effective central charges.
We follow closely the cases without boundary considered in \cite{AFKRY}.
Comparison of these results with numerical solutions of the TBA equations
and with the vacuum energies are presented in sect.5.
We conclude in sect.6 with open questions and some remarks.

%%%%%%%%%%%%%%%%%%%%%%%%%%%%%%%%%%%%%%%%%%%%%%%%%%%%%%%%%%%%%%%%%

\section{Affine Toda Theories on a half line}

\subsection{Integrable Actions, Bulk and Boundary $S$-Matrices}

The ATFTs corresponding to Lie algebra  $G$ is 
described by the action
\beq
{\cal A}=\int d^2x\left[{1\over{8\pi}}(\pd_{a}\vph)^2
+\mu\sum_{i=0}^{r}e^{b\bfe_i\cdot\vph}\right]
+\mu_B\int dx\sum_{i=0}^{r}\eta_i e^{b\bfe_i\cdot\vph/2},
\label{action}
\eeq
where $\bfe_i,\ i=1,\ldots,r$ are the simple roots of the Lie algebra
$G$ of rank $r$ and $-\bfe_0$ is a maximal root satisfying
\beq
\bfe_0+\sum_{i=1}^{r}n_i\bfe_i=0.
\label{maxroot}
\eeq

For real $b$ the spectrum of these ATFTs consists of $r$ particles 
with the masses $m_i$ ($i=1,\ldots,r$) given by
\beq
m_i={\overline m}\nu_i
\eeq
where 
\beq
{\overline m}^2={1\over{2h}}\sum_{i=1}^{r}m_i^2
\eeq
and here $h$ is Coxeter number and $\nu_i^2$ are the eigenvalues of 
the mass matrix:
\beq
M_{ab}=\sum_{i=0}^{r}(\bfe_i)^a(\bfe_i)^b.
\eeq

The mass-$\mu$ relation is
\beq
-\pi\mu\ga(1+b^2)=\left[{{\overline m}k(G)\Ga\left({1\over{
(1+b^2)h}}\right)\Ga\left(1+{b^2\over{(1+b^2)h}}\right)\over{
2\Ga(1/h)}}\right]^{2(1+b^2)}
\label{mmu}
\eeq
where $\ga(x)=\Ga(x)/\Ga(1-x)$ and
\beq
k(G)=\left(\prod_{i=1}^{r}n_i^{n_i}\right)^{1/2h}
\label{kofg}
\eeq
with $n_i$ defined in Eq.(\ref{maxroot}).

The scattering amplitudes of the ATFTs are
factorized into the two-particle bulk $S$-matrices.
From the bootstrap relations, crossing symmetry, and unitarity,
the $S$-matrix between the particles $m_i$ and $m_j$ are given by
\cite{AFZ,BCDS}:
\beaq
S_{ij}(\th)&=&\exp(-i\delta_{ij}(\th))\qquad{\rm where}
\label{Satft}\\
\delta_{ij}(\th)&=&\int_{0}^{\infty}
{dt\over{t}}\left[8\sinh\left({\pi B t\over{h}}\right)
\sinh\left({\pi(1-B)t\over{h}}\right)
{\left(2\cosh{\pi t\over{h}}-{\bf I}\right)^{-1}}_{ij}
-2\de_{ij}\right]\sin(\th t),\nonumber
\eeaq
where ${\bf I}$ is the incident matrix defined by
${\bf I}_{ij}=2\de_{ij}-\bfe_i\cdot\bfe_j$ and with
\beq
B={b^2\over{1+b^2}}.
\label{coupling}
\eeq

The second term in (\ref{action}) is the boundary action which 
preserves the integrability. 
The boundary parameter $\mu_B$ should be fixed completely to have
conserved charges \cite{CDRS,FZZ} 
\beq
\mu_B^2={\mu\over{2}}\cot\left({\pi b^2\over{2}}\right)
\label{NATTbs}
\eeq
with discrete parameter 
\beq
\eta_i=1,\quad -1,\quad 0.\label{discrete}
\eeq
Only exception is the $A_1$-ATFT, namely the sinh-Gordon
model, which include two continuous free parameters on boundary.
These extra parameters introduce additional complicacy in the analysis
\cite{Alyosha} and will not be considered here.

With the integrable boundary actions, the boundary $S$-matrices
should satisfy the boundary Yang-Baxter equations and boundary 
bootstrap relations \cite{fring,sasaki} along with boundary 
crossing-unitarity relation \cite{GhoZam}.
With diagonal bulk $S$-matrices of the ATFTs, these relations can
determine the boundary $S$-matrices only upto CDD factors.
This CDD ambiguity is more serious for the IQFTs with boundary.
While the bulk $S$-matrices can be checked both 
perturbatively and nonperturbatively, the boundary $S$-matrices 
is difficult to check perturbatively due to the presence of boundary 
\cite{Kim}, not to mention non-perturbative checks.
In addition, there is no clear relations between the boundary actions
and specific CDD factors to be chosen.
So far, only a few boundary $S$-matrices have been associated with 
combinations of the discrete possibilities Eq.(\ref{discrete}).
In this paper we will consider only the simplest case where all
the parameters $\eta_i=+1$ denoted by $(+)$ boundary condition (BC)
and the Neumann (free) BC denoted by $(f)$ with all $\eta_i=0$,
which is conjectured to be a dual ($b\to 1/b$) of $(+)$ \cite{bdual}.
We will impose the two BCs $(+)$ and $(f)$ independently on both 
boundaries of the strip.
These are the cases without boundary bound states where the 
standard ``ground state'' TBA should work.

For $A_{n-1}^{(1)}$ ATFTs, 
the boundary $S$-matrices $R_{j}(\th)$ for the $(+)$ BCs are 
given by \cite{CDRS}
\beq
R^{(+)}_{j}(b,\th)=\prod_{a=1}^{j}[a-1][a-n][-a+B][-a-n+1-B],\quad
j=1,\ldots,n-1 \label{aap}
\eeq
where
\beq
[x]={\sinh(\th/2+i\pi x/2h)\over{\sinh(\th/2-i\pi x/2h)}}.
\label{bracket}
\eeq
The boundary $S$-matrices for the free BCs are conjectured to be
dual transform of the $(+)$ BCs, namely,
\beq
R^{(f)}_{j}(b,\th)=R^{(+)}_{j}(1/b,\th).
\eeq
We will consider only the $A$-type ATFTs to avoid extra complicacy
arising from ambiguous CDD factors for $D$- and $E$-type ATFTs. 

\subsection{Boundary TBA}

TBA equations are constructed in the rectangle with each size $R$ and $L$.
The free energy given by scattering theories defined on the space 
with infinite size $L$ and imaginary time $R=1/T$ (``$R$-channel'')
is compared with the opposite case where the Casimir energy $E_{0}(R)$ 
is obtained as a function of finite spatial size $R$ while the time $L$ 
(``$L$-channel'') goes to $\infty$.
With periodic BCs, the opposite case of $L$ finite and
$R\to\infty$ does not raise any new problem since it is identical to the
above by just interchanging $R$ and $L$.  

In the presence of boundaries where specific integrable BCs
are imposed, these two cases have totally different meanings because
one of the size, say $R$, should denote the width between two sides
where the BCs are imposed \cite{LMSS}.
Now consider the former case.
In the $R$-channel, the thermodynamic functions are defined 
by the scattering theories with ``fugacity'' where the boundary states 
act as creating/annihilating sources of particle pairs with certain 
probabilities which are determined by the boundary $S$-matrices.
The physical quantity generated by the TBA is the Casimir energy
$E_{0}^{\al\be}(R)$ which depends on the specific BCs
$(\al)$ and $(\be)$.
The Casimir energy is related to the effective central charges
of the underlying CFTs.
In opposite case of $R\to\infty$, the thermodynamic analysis
generates the boundary entropy instead.

For the irrational CFTs like the NATTs, it is quite difficult
to define conformal boundary states and associated boundary entropies.
Therefore, we will concentrate on the former case of $L\to\infty$
with finite $R$ in which the boundary TBA generates the effective
central charge.

Following the the formalism of \cite{LMSS}, we can derive the TBA
equations straightforwardly because the boundary ATFTs
are purely diagonal scattering theories.
The TBA equations for the ATFTs are given by ($i=1, \cdots, r$)
\beq\label{Btba}
2m_i R \cosh\theta=\ep_i (\theta)+
\sum_{j=1}^r\int_{-\infty}^{\infty}\varphi_{ij} (\theta-\theta')
\log\left(1+\la^{(\al\be)}_{i}(\th') 
e^{-\ep_i(\th')}\right){d\theta'\over 2\pi},
\eeq
where $\varphi_{ij}$ is the kernel which is equal to the logarithmic
derivative of the $S$-matrix $S_{ij} (\th)$ in Eq.(\ref{Satft})
\[
\varphi_{ij} (\th)=-i {d \over d\th}\log S_{ij} (\th)=\de'_{ij}(\th)
\]
and 
\beq
\la^{(\al\be)}_{j}(\th)=R^{(\al)}_j\left(\th+{i\pi\over{2}}\right)
R^{(\be)}_j\left(-\th+{i\pi\over{2}}\right)
\eeq
where $(\al)$ and $(\be)$ refer to the integrable BCs
either $(+)$ or $(f)$.

The `pseudo-energies' $\ep_i(\theta,R)$ give the scaling function of
the effective central charge
\beq  
\label{ceff_tba}
c_{\rm eff}(R) = \sum_{i=1}^r\frac{6Rm_i}{\pi^2}
\int \cosh\theta \log\left(1+\la^{(\al\be)}_{i}(\th) 
e^{-\ep_i(\th)}\right)d\theta.
\eeq

We compute the effective central charges for the
simply-laced $A_n$ ATFTs with the BCs on both boundaries
$(++)$, $(+f)$, $(f+)$, and $(ff)$
and compare with the UV
asymptotics determined by the reflection amplitudes which will be
derived in the next section. 
This provides nonperturbative check for the boundary
$S$-matrices conjectured in the literature.

\section{Reflections of Quantum Mechanical Waves}
In this section we will follow the same logical step as the NATTs without
boundary in \cite{AFKRY} to derive the ``boundary reflection
amplitudes'' of the boundary NATTs. 

\subsection{boundary Liouville theory}
We start with the LFT with boundary whose action is given by 
\begin{equation}
A_{\mathrm{Liouv}}=\int\limits_{y_1 \le y \le y_2 }
\left(  \frac{1}{4\pi}(\partial_{a}%
\phi)^{2}+ \mu e^{2b\phi }\right)  d^{2}x
+\mu_{B}^{(1)} \int e^{b\phi_B (y_1) } dx 
+\mu_{B}^{(2)} \int e^{b\phi_B (y_2) } dx\,.
\end{equation}

The ``boundary reflection amplitude'' of the LFT relating a conjugate
pair of boundary operators $V_{\be}\equiv\exp{\be\varphi_{\rm B}}$ and
$V_{Q-\be}$ of the same dimension are defined by
\beq
\langle V_{\be}(x)V_{\al_1}(x_1)\ldots\rangle=d(\be|s_1,s_2)
\langle V_{Q-\be}(x)V_{\al_1}(x_1)\ldots\rangle,
\eeq
where the parameter $s$ is given by
\beq
\cosh^2(\pi bs)={\mu_{B}^2\over{\mu}}\sin(\pi b^2).
\label{LFTbs}
\eeq 
It is more convenient to use a real variable $P$ defined by
$\be=Q/2+iP$ and define 
\beq
S_B(P|s_{1},s_{2})=d(Q/2+iP|s_{1},s_{2})
\eeq
since the reflection $\be\to Q-\be$ corresponds to $P\to
-P$ in this parametrization so that the ``reflection'' has a physical
meaning in this parameter space.
This quantity has been obtained by functional relations and boundary 
degenerate operators in \cite{FZZ} as follows:
\beq
S_B(P|s_{1},s_{2})=
\left(
\pi\mu\gamma(b^{2})b^{2-2b^{2}}\right)^{-iP/b}{G_B(-P|s_{1},s_{2})
\over{G_B(P|s_{1},s_{2})}}
\label{bLFTref}
\eeq
where
\bea
G_B(P|s_{1},s_{2})&=& G (2iP)
\frac{G(Q/2 -iP -i(s_1 -s_2)/2) G(Q/2 -iP +i(s_1 -s_2)/2) }
{G(Q/2 +iP -i(s_1 + s_2)/2) G(Q/2 -iP +i(s_1 + s_2)/2) } \,.
\eea
Here the function $G(x)$ is explicitly defined as
\beq
\log G(x)=
\int\limits_{0}^{\infty}
\frac{dt}t\left[\frac
{e^{-Qt/2}-e^{-xt}}{(1-e^{-bt})(1-e^{-t/b})}+\frac{(Q/2-x)^{2}}2 e^{-t}%
+ \frac{Q/2 - x}{t}\right]\,.
\eeq

One can expand the scalar field in terms of zero-mode and oscillator modes
\beq 
\varphi(x)=\varphi_0-{\cal P}(z-{\overline z})+ \sum_{n\neq 0}
\left({i a_n\over{n}}e^{inz}+ {i{\overline a}_n\over{n}}
e^{-in{\overline z}}\right)
\eeq 
in the limit of $\varphi_0\to-\infty$ where the interaction terms vanish.
Boundary condition imposes a constraint $a_n={\overline a}_n$. 
Then, a primary field $V_{\be}$ can be described
in the $b\to 0$ limit by a wave functional
\beq
\Psi_{P}[\varphi(x)]\sim \Psi_{P}(\varphi_0)\otimes\vert 0\rangle
\eeq 
satisfying the Liouville Schr\"odinger equation:
\beq 
\left[{1\over{24}}-\del_{\varphi_0}^2 +\pi\mu
e^{2b \varphi_0}+\mu_{B}e^{b\varphi_0} \right] \Psi_{P}(\varphi_0) =
E_0\Psi_{P}(\varphi_0) 
\eeq
where $E_0$ is the ground-state energy 
\beq
E_0=-{1\over{24}}+P^2 
\eeq
and $\mu_B = \mu_B^{(1)} + \mu_B^{(2)}$.

This interpretation of the ``boundary reflection amplitude'' as
quantum mechanical amplitude of the zero-mode wave
function reflected off from the exponential potential wall
can be confirmed by solving the Schr\"odinger equation. 
The solution is written in terms of confluent hypergeometric function
\beq 
\Psi_{P}(\varphi_0) =  N z^\nu e^{-z/2}
\left[ { M(C,D,z ) \over \Gamma(1+C -D) \Gamma (D) } -
z^{1-D} { M(1+C-D,2-D,z ) \over \Gamma(C ) \Gamma (2-D) }
\right],
\eeq
where $N$ is a normalization constant,
$M(C,D,z )= {}_1F_1(C;D;z) $ is the Kummer function and
\[
z = {2\sqrt{ \pi \mu  }\over{b}} e^{b\phi_0 },\quad
C ={1\over{2}} +{\mu_B\over{2\sqrt{\pi \mu b^2 }}}+{iP\over{b}},\quad 
D = 1 + {2i P\over{b}}.
\]
In the limit of $\varphi_0\to-\infty$, the wave functional behaves like 
\beq 
\Psi_{P}(\varphi_0)\sim
e^{iP\varphi_0}+{\tilde S}_B(P)e^{-iP\varphi_0} 
\eeq
where
\beq
{\tilde S}_B(P) = \left( {4\pi \mu  \over b^2}\right)^{-iP/b}
\frac{\Gamma (\frac12 + \frac{\mu_B }{2 b \sqrt{\pi\mu}}
-\frac{iP}{b})
\Gamma(\frac{2iP}{b}) }
{\Gamma (\frac12 + \frac{\mu_B }{2 b \sqrt{\pi\mu}}
+\frac{iP}{b}))
\Gamma(- \frac{2iP}{b}) }.
\label{RefAmp}
\eeq
This ${\tilde S}_B(P)$ indeed reproduces the reflection amplitude
$S_B(P|s_{1},s_{2})$ in Eq.(\ref{bLFTref}) as $b\to 0$.
For generic value of $b$ we will use $S_B(P|s_{1},s_{2})$ as the
quantum mechanical reflection amplitude.

\subsection{Non-affine Toda theories}

Now we generalize the result on the boundary LFT to the NATTs.
The actions of these models can be obtained by removing the affine
terms associated with $\bfe_0$ from those of the ATFTs (\ref{action}).
In the presence of boundary, the primary fields of the NATTs can be 
described by the wave functionals $\Psi[\vph(x)]$
whose asymptotic behaviours are described by the wave functions of
the zero-modes.
The zero-modes of the fields $\vph(x)$ are defined as:
\beq
\vph_0 = \int_0^{\pi}\vph(x) \frac{dy}{\pi}.
\eeq
Here we consider the NATT on an infinite strip of width
$\pi$ with coordinate $x$ along the strip playing the role
of imaginary time.
In the asymptotic region where the potential terms in the NATT action 
become negligible ($\bfe_{i}\cdot\vph_0\to-\infty$ for all $i$), 
the fields can be expanded in terms of free field operators $\va_n$
\beq
\vph(x)=\vph_0-{\cal P}(z-{\overline z})+
\sum_{n\neq 0}\left({i\va_n\over{n}}e^{inz}+
{i{\overline\va}_n\over{n}}e^{-in{\overline z}}\right),
\label{free}
\eeq
where ${\cal P}=-i\del_{\vph_0}$ is the conjugate momentum of $\vph_0$.

In this region any state of the NATT can be decomposed into a direct 
product of two parts, namely, a wave function of the zero-modes and a state 
in Fock space generated by the operators $\va_n$.
The physical states should satisfy the constraint equations
\beq
(\va_n-{\overline\va}_n)|s\rangle=0.
\eeq
In particular, the wave functional corresponding to the primary state
can be expressed as a direct product of a wave function 
of the zero-modes $\vph_0$ and Fock vacuum: 
\beq
\Psi_{\vP}[\vph(x)]\sim \Psi_{\vP}(\vph_0)\otimes\vert 0\rangle
\label{vip}
\eeq
where the wave function $\Psi_{\vP}(\vph_0)$ in this asymptotic region
is a superposition of plane waves with momenta ${\hat s}\vP$.

The reflection amplitudes of the NATT defined in the previous section
can be interpreted as those for the wave function of the zero-modes
in the presence of potential walls.
This can be understood most clearly in the semiclassical limit $b\to 0$
where one can neglect the operators $\va_n$ 
even for significant values of the parameter $\mu$.
The full quantum effect can be implemented simply by introducing
the exact reflection amplitudes which take into account also non-zero-mode
contributions.
The resulting Schr\"odinger equation is given by
\beq
\left[{r\over{24}}-2\del_{\vph_0}^2
+\mu\pi\sum_{i=1}^{r}e^{b\bfe_i\cdot\vph_0}
+\mu_{B}\sum_{i=1}^{r}A_i e^{b\bfe_i\cdot\vph_0/2}\right]
\Psi_{\vP}(\vph_0) = E_0\Psi_{\vP}(\vph_0)
\eeq
with the ground state energy
\beq
E_0=-{r\over{24}}+2\vP^2.
\label{energy}
\eeq
Here the momentum $\vP$ is any continuous real vector.
The effective central charge can be obtained from Eq.(\ref{energy})
where $\vP^2$ takes the minimal possible value for the perturbed theory.
Since only asymptotic form of the wave function matters, we can derive
the reflection amplitudes in the same way as the ATFTs without
boundary \cite{AFKRY}.

In the UV limit where $\mu, \mu_B \rightarrow 0$,
the potential vanishes almost everywhere 
except for the values of $\vph_0$ where some of exponential terms in the 
potential become large enough to overcome the small value of $\mu$. 
In this case, each exponential term
$e^{b \bfe_i\cdot\vph_0}$ in the interaction represent a wall 
with $\bfe_i$ being its normal vector. 
If we consider the behaviour of a wave function near a wall normal to
$\bfe_i$ where the effect of other interaction terms becomes negligible, 
the problem becomes equivalent to the boundary LFT in the $\bfe_i$ direction.
The potential becomes flat in the $(r-1)$-dimensional orthogonal 
directions.
The asymptotic form of the energy eigenfunction is then given by the 
product of that of Liouville wave function and $(r-1)$-dimensional plane wave,
\beaq \label{nearwall}
\Psi_{\vP} &\sim& \left[ e^{i P_i\vp_{0i}} + S_B(P_i) 
e^{-i P_i\vp_{0i}} \right] e^{i\vP_\perp\cdot\vph_0} \nonumber\\
&\sim& e^{i \vP\cdot\vph_0} + S_B(P_i)e^{i \hat{s}_i\vP\cdot\vph_0}\,,
\eeaq
where $\hat{s}_i$ denotes the Weyl reflection by the simple root $\bfe_i$
and $P_i$ the component of $\vP$ along $\bfe_i$ direction.

We can see from Eq.(\ref{nearwall}) that the momentum of the
reflected wave by the $i$-th wall is given by the Weyl reflection
$\hat{s}_i$ acting on the incoming momentum.
If we consider the reflections from all the potential walls,
the wave function in the asymptotic region is a superposition of the
plane waves reflected by potential walls in different ways.
The momenta of these waves form the orbit of the Weyl group
${\cal W}$ of the Lie algebra $G$;
\beq \label{wavefunction}
\Psi_\vP(\vph_0)\simeq\sum_{\hat{s}\in {\cal W}}
A(\hat{s}\vP)e^{i\hat{s}\vP\cdot\vph_0}.
\eeq
This is indeed the wave function representation of the primary field
in the asymptotic region.

It follows from Eq.(\ref{nearwall}) that the amplitudes $A(\vP)$ satisfy 
the relations
\beq \label{ratio}
\frac{A(\hat{s}_i\vP)}{A(\vP)} = S_B(P_i).
\eeq
For a general Weyl element ${\hat s}$ which can be represented by a product 
of the Weyl elements ${\hat s}_i$ associated with the simple roots
by $\hat{s} = \hat{s}_{i_k}\hat{s}_{i_{k-1}}\cdots\hat{s}_{i_1}$,
the above equation can be generalized to
\beq
\frac{A(\hat{s}_{i_k}\cdots\hat{s}_{i_1}\vP)}{A(\vP)}
= S_B(\vP\cdot\bfe_{i_1})S_B(\hat{s}_{i_1}\vP\cdot\bfe_{i_2})
S_B(\hat{s}_{i_{2}}\hat{s}_{i_1}\vP\cdot\bfe_{i_3})
\cdots S_B(\hat{s}_{i_{k-1}}\cdots\hat{s}_{i_1}\vP\cdot\bfe_{i_k}).
\label{genref}
\eeq
Using the properties of the Weyl group 
and the explicit form of the amplitude $S_{B}(P)$ in Eq.(\ref{bLFTref}),
it is straightforward to verify that the following function $A(\vP)$ 
satisfies Eqs.(\ref{ratio}) and (\ref{genref}):
\beq \label{amplitude}
A(\vP) = \left(\pi\mu\gamma(b^2)b^{2-2b^2}\right)^{i\vrh\cdot\vP/b}
\prod_{\val>0}G_B(P_{\val}|s_1,s_2)
\eeq
where $P_{\val}= \val\cdot\vP$ is a scalar product with a positive root 
$\val$.  
This result is valid for all simply-laced Lie algebras.
Major difference in the NATTs is that the values of $s$ parameters
should be fixed since there are no free boundary parameters in the
NATTs. 
Comparing Eqs.(\ref{NATTbs}) and (\ref{LFTbs}), one can find that
for the $(+)$ BC
\beq
s^{(+)}={ib\over{2}},
\label{param}
\eeq
and for the free BC $(f)$ using duality
\beq
s^{(f)}={i\over{2b}}.
\label{paramdual}
\eeq
Therefore, the  boundary reflection amplitudes of the NATTs for the 
four combinations of the two BCs are given by 
Eq.(\ref{bLFTref}) with $s_1$ and $s_2$ 
\bea
(++)\quad&\leftrightarrow&\quad s_1={ib\over{2}},\quad s_2={ib\over{2}}\\
(+f)\quad&\leftrightarrow&\quad s_1={ib\over{2}},\quad s_2={i\over{2b}}\\
(f+)\quad&\leftrightarrow&\quad s_1={i\over{2b}},\quad s_2={ib\over{2}}\\
(ff)\quad&\leftrightarrow&\quad s_1={i\over{2b}},\quad s_2={i\over{2b}}.
\eea

The boundary reflection amplitudes of the
NATTs with $(++)$ BCs have been derived more rigorously 
from the generalized functional relations method along with 
boundary degenerate operators in \cite{Fateev}.
It is straightforward to check that these two independent derivations
match exactly.
This confirms the validity of our wave functional interpretation.

\section{Quantization Condition and Scaling Function}

In this section we derive the UV asymptotic expressions for the
effective central charges using the quantization conditions satisfied
by the wave funcionals confined in the potential well. 
When perturbed by the bulk and boundary operators associated with the
affine root, the NATTs become the ATFTs with boundary Eq.(\ref{action}). 
The perturbations provide additional potential wall which confines
the wave functional in the multi-dimensional potential well,
i.e. the Weyl chamber.
Once confined in the well, the wave functional is quantized and has
discrete energy levels. 
The derivation of the quantization condition is exactly  
identical to the bulk only case \cite{AFKRY} if one substitute the
bulk reflection amplitude to the boundary one.
The quantization condition becomes
\beq \label{quant2}
\left(\pi\mu\gamma(b^2)b^{2-2b^2}\right)^{ih\vP\cdot\hat{s}\bfe_0/b}
\prod_{\val>0}\left[\frac{G_B(-\vP\cdot\hat{s}\val)}
{G_B(\vP\cdot\hat{s}\val)}
\right]^{-\hat{s}\val\cdot\hat{s}\bfe_0} = 1.
\eeq
Since the Weyl element $\hat{s}$ is arbitrary, Eq.(\ref{quant2}) 
leads to the following condition for the lowest energy state 
\beq \label{quantization}
2hQL\vP = 2\pi\vrh - \sum_{\val>0}\val\de_{B}(P_{\val})\,,
\eeq
where
\[
L=-\frac{1}{2(1+b^2)}\log[\pi\mu\gamma(b^2)b^{2-2b^2}]\,,
\]
and
\beq \label{delta}
\de_B(P) = i\log{G_B(P)\over{G_B(-P)}}.
\eeq
This is the quantization condition for the momentum $\vP$ in the
$\mu\rightarrow 0$ limit. 
We see that each positive root $\val$ causes an effective 
phase shift of Liouville type.

Now we consider the system defined on a strip with a width $R$.
When we scale back the size from $R$ to $\pi$,
the parameter $\mu$ in the action (\ref{action}) changes to
\beq	
\mu \rightarrow \mu \left(\frac{R}{\pi}\right)^{2(1+b^2)}.
\eeq
The $\mu\rightarrow 0$ limit is realized as the deep UV limit $R\to 0$.
The rescaling changes the definition of $L$ in Eq.(\ref{quantization}) by
\beq \label{L}
L=-\log\frac{R}{\pi}-\frac{1}{2(1+b^2)}
\log[\pi\mu\gamma(b^2)b^{2-2b^2}]\,.
\eeq
The ground state energy with the circumference $R$ is given by
\beq
E(R)=-{\pi c_{\rm eff}\over{24R}}\quad
{\rm with}\quad
c_{\rm eff}=r-48\vP^2
\label{grenergy}
\eeq
where $\vP$ satisfies Eq.(\ref{quantization}).

In this limit, Eq.(\ref{quantization}) can be solved perturbatively. 
For this we expand the function $\de_B(P)$ in Eq.(\ref{delta}) 
in powers of $P$,
\beq \label{expand}
\de_B(P) = \de_1(b)P + \de_3(b)P^3 + \de_5(b) P^5 \cdots\,,
\eeq
where
\beaq \label{deltas}
\de_1(b) &=& -2\gamma_{E}(b+1/b)-2(b-1/b)\log b \nonumber\\
& &+2\int_{0}^{\infty}{dt\over{t}}\left[{2t\cosh^2(bt/2)\over{
\sinh(bt)\sinh(t/b)}}-{t(e^{-bt}+e^{-t/b})\over{(1-e^{-bt})(1-e^{-t/b})}}
\right]\nonumber\\
\de_3(b) &=& {8\over{3}}(b^3+b^{-3})\zeta(3)-{8\over{3}}
\int_{0}^{\infty}dt\left[{t^2\cosh^2(bt/2)\over{
\sinh(bt)\sinh(t/b)}}-{t^2 (e^{-bt}+e^{-t/b})\over{2(1-e^{-bt})(1-e^{-t/b})}}
\right]\nonumber\\
\de_5(b) &=& -{32\over{5}}(b^5+b^{-5})\zeta(5)
+{8\over{15}}\int_{0}^{\infty}dt
\left[{t^4\cosh^2(bt/2)\over{
\sinh(bt)\sinh(t/b)}}-{t^4(e^{-bt}+e^{-t/b})\over{2(1-e^{-bt})(1-e^{-t/b})}}
\right].\nonumber
\eeaq
Using the relation 
\[
\sum_{\val>0} (\val)^a(\val)^b = h\de^{ab},
\] 
we obtain
\[
hl\vP = 2\pi\vrh-\de_3(b)\sum_{\val>0}\val(P_{\val})^3
-\de_5(b)\sum_{\val>0}\val(P_{\val})^5 - \cdots\,,
\]
with
\beq \label{smalll}
l\equiv 2QL + \de_1\,.
\eeq
The above equation can be solved iteratively in powers of $1/l$. 
Inserting the solution into Eq.(\ref{grenergy}), we find
\beq
c_{\rm eff}=r-\frac{4r(h+1)}{h}\left[\left(\frac{2\pi}{l}\right)^2
-\frac{24\delta _3}{2\pi}\left(\frac{2\pi}{l}\right)^5D_4
-\frac{24\delta _5}{2\pi}\left(\frac{2\pi}{l}\right)^7D_6
+{\cal O}(l^{-8})
\right]
\label{ceff}
\eeq
where the coefficients are given by
\[
D_4=\frac{1}{r(h+1)h^4}\sum_{\val >0}\rho_{\val}^4=
{2n^2-3\over{60n^2}},\qquad
D_6=\frac{1}{r(h+1)h^6}\sum_{\val >0}\rho_{\val}^6=
{(n^2-2)(3n^2-5)\over{168n^4}},
\]
for the $A_{n-1}$ algebra.

\section{Numerical Comparison}
It is quite difficult task to solve the TBA equations analytically and
compare directly with Eq.(\ref{ceff}).
To obtain higher order terms in $1/l$ expansion, one needs to solve
complicated coupled nonlinear differential equations.
Even the lowest order terms at the order of $1/l^2$ contain constants
which can not be decided by the scattering data.
Even the numerical analysis is not easy when the rank $r$ grows
because a large number of equations amplify the numerical errors
entering in the iteration procedure.
We will consider $A_2$, $A_3$, and $A_4$ ATFTs and compute  
effective central charges $c_{\rm eff}(R)$ by solving Eq.(\ref{ceff_tba}) 
iteratively as a function of ${\overline m}R$ for the BCs 
mentioned above.
In order to compare the numerical data with our results based on
the reflection amplitudes, 
we fit the numerical data for $c_{\rm eff}(R)$ from the TBA equations 
for many different values of $R$ with the function (\ref{ceff}) where
$\de_1$, $\de_3$ and $\de_5$ are considered as the fitting parameters. 
For this comparison the relation
(\ref{mmu}) between the parameter $\mu$ in the action and parameter
${\overline m}$ for the particle masses is used.
These parameters $\de_i$'s are then compared with Eq.(\ref{deltas}) 
defined from the reflection amplitude of the LFT. 
Since we already separate out the dependence on the Lie algebra $G$,
our numerical results for the parameters $\de_i$'s should be 
independent of $G$.

\subsection{$(++)$ and $(ff)$ boundary conditions}

These two BCs are related by the dual transform $B\to 1-B$.
Hence, it is enough to consider $(++)$ BC only for $0<B<1$. 
The fugacity is given by
\beq
\la^{(++)}_{j}(\th)=R^{(+)}_j\left(\th+{i\pi\over{2}}\right)
R^{(+)}_j\left(-\th+{i\pi\over{2}}\right)
\eeq
and the reflection amplitudes are given by
Eq.(\ref{quantization}) with $s_1=s_2=ib/2$.
Tables 1--3 show the values of parameters $\de_i$'s obtained numerically 
from TBA equations for various values of the coupling constant $B$ 
in $A_2$, $A_3$, and $A_4$ ATFTs. 
We see that they are in excellent agreement with those values of $\de_i$'s 
following from the reflection amplitudes supplemented with Eq.(\ref{mmu}).
Thus numerical TBA analysis fully supports the validity of our whole scheme
based on the reflection amplitude, $\mu$-${\overline m}$ relation, the
shift and the quantization condition on $\vP$. 
\begin{table}
\centerline{
\begin{tabular}{||c||c||c|c|c||} \hline
\rule[-.4cm]{0cm}{1.cm}B
& $\de^{\rm (RA)}_1$ & $\de_1^{\rm (TBA)}(A_2)$
& $\de_1^{\rm (TBA)}(A_3)$ & $\de_1^{\rm (TBA)}(A_4)$ \\ \hline
0.25 & --8.12340 & --8.12344 & --8.12345 & --8.12345 \\
0.30 & --6.15569 & --6.15571 & --6.15572 & --6.15572 \\
0.35 & --4.73422 & --4.73423 & --4.73423 & --4.73423 \\
0.40 & --3.68012 & --3.68013 & --3.68013 & --3.68013 \\
0.45 & --2.89175 & --2.89175 & --2.89175 & --2.89175 \\
0.50 & --2.30886 & --2.30886 & --2.30886 & --2.30887 \\
0.55 & --1.89627 & --1.89627 & --1.89627 & --1.89627 \\
0.60 & --1.63628 & --1.63628 & --1.63628 & --1.63628 \\
0.65 & --1.52627 & --1.52627 & --1.52627 & --1.52627 \\
0.70 & --1.58079 & --1.58079 & --1.58079 & --1.58079 \\
0.75 & --1.84022 & --1.84022 & --1.84022 & --1.84022 \\
0.80 & --2.39472 & --2.39472 & --2.39473 & --2.39455 \\ \hline
\end{tabular}
}
\caption{$\de_1^{\rm (RA)}$ vs.
$\de_1^{\rm (TBA)}$ for $A_2,A_3$, and $A_4$ ATFTs for $(++)$ BC.}
\end{table}
\begin{table}[t]
\centerline{
\begin{tabular}{||c||c||c|c|c||} \hline
\rule[-.4cm]{0cm}{1.cm}B
& $\de^{\rm (RA)}_3$ & $\de_3^{\rm (TBA)}(A_2)$
& $\de_3^{\rm (TBA)}(A_3)$ & $\de_3^{\rm (TBA)}(A_4)$ \\ \hline
0.25 & 32.5848 & 32.6563 & 32.6634 & 32.6677 \\
0.30 & 22.1701 & 22.2054 & 22.2095 & 22.2113 \\
0.35 & 15.6353 & 15.6524 & 15.6549 & 15.6560 \\
0.40 & 11.3251 & 11.3327 & 11.3343 & 11.3350 \\
0.45 & 8.40246 & 8.40505 & 8.40615 & 8.40663 \\
0.50 & 6.41097 & 6.41084 & 6.41165 & 6.41200 \\
0.55 & 5.09233 & 5.09070 & 5.09136 & 5.09164 \\
0.60 & 4.30543 & 4.30291 & 4.30349 & 4.30374 \\
0.65 & 3.99324 & 3.99008 & 3.99064 & 3.99088 \\
0.70 & 4.18219 & 4.17842 & 4.17903 & 4.17928 \\
0.75 & 5.02369 & 5.01914 & 5.01987 & 5.02180 \\
0.80 & 6.93616 & 6.93039 & 6.93375 & 7.29485 \\ \hline
\end{tabular}
}
\caption{$\de_3^{\rm (RA)}$ vs.
$\de_3^{\rm (TBA)}$ for $A_2,A_3$, and $A_4$ ATFTs for $(++)$ BC.}
\end{table}
\begin{table}[t]
\centerline{
\begin{tabular}{||c||c||c|c|c||} \hline
\rule[-.4cm]{0cm}{1.cm}B
& $\de^{\rm (RA)}_5$ & $\de_5^{\rm (TBA)}(A_2)$
& $\de_5^{\rm (TBA)}(A_3)$ & $\de_5^{\rm (TBA)}(A_4)$ \\ \hline
0.25 & --206.638 & --208.099 & --208.050 & --208.047 \\
0.30 & --110.087 & --110.739 & --110.726 & --110.720 \\
0.35 & --62.0830 & --62.3780 & --62.3798 & --62.3796 \\
0.40 & --36.3095 & --36.4358 & --36.4434 & --36.4458 \\
0.45 & --21.7458 & --21.7885 & --21.7986 & --21.8021 \\
0.50 & --13.2727 & --13.2721 & --13.2833 & --13.2874 \\
0.55 & --8.33556 & --8.31104 & --8.32307 & --8.32744 \\
0.60 & --5.61763 & --5.57761 & --5.59071 & --5.59551 \\
0.65 & --4.48789 & --4.43462 & --4.44955 & --4.45501 \\
0.70 & --4.82305 & --4.75409 & --4.77232 & --4.77893 \\
0.75 & --7.15865 & --7.06574 & --7.09021 & --7.12790 \\
0.80 & --13.5875 & --13.4502 & --13.5398 & --21.0323 \\ \hline
\end{tabular}
}
\caption{$\de_5^{\rm (RA)}$ vs.
$\de_5^{\rm (TBA)}$ for $A_2,A_3$, and $A_4$ ATFTs for $(++)$ BC.}
\end{table}

The agreement of $\de_5$ becomes less accurate for the cases with 
high rank partly due to the numerical errors in higher order calculations. 
Another reason comes from the fact that neglected terms in the 
$1/l$ expansion (the order of ${\cal O}(1/l^8)$ or higher) in 
Eq.(\ref{ceff}) may not be sufficiently small 
compared with terms with $\de_5$. 
However, one can in principle reduce these errors by
increasing the accuracy of the numerical calculations.
There are also corrections of ${\cal O}(R^{\gamma})$ to the expansion
of $c_{\rm eff}(R)$ in power series of $1/l$ which increase
as $B$ goes to zero. 
This explains why the discrepancies in the tables increase as $B$ decreases.

In Fig.1, we also plot the scaling functions $c_{\rm eff}(R)$ as a 
function of $R$ setting ${\overline m}=1$ for different ATFTs in two
ways.
The first is the curves generated by the TBA equations and the second
by the reflection amplitudes. 
To compare the same objects, one should add to the second case the 
contribution from the vacuum energy terms.
In addition to the dual symmetric bulk contribution given by \cite{DDV}
\beq
\Delta c_{\rm Bulk}={3{\overline m}^2R^2\over{2\pi}}
{\sin(\pi/h)\over{\sin(\pi B/h)\sin(\pi(1-B)/h)}},
\eeq
one should consider the boundary contribution. 
This is given by $2\Delta c_{\rm boun}$ where
$\Delta c_{\rm boun}$ is the boundary vacuum energy obtained in 
\cite{Fateev} 
\beq
\Delta c_{\rm boun}(b)={6{\overline m}R
\cos(\pi/2h)\over{\sin(\pi B/2h)\sin(\pi(1-B)/2h)}}.
\eeq
These terms are negligible in the UV region.
The ``experimental'' observation that two results agree well
even for large values of $R$ can provide nonperturbative check for the
boundary vacuum energies.
%%%%%%%%%%%%%%%
To illustrate the accurate agreement, we plot $c_{\rm eff}(R)$ as a
function of $R$ for $A_2$ ATFT in Fig.2.
The dotted line is without any vacuum energies. 
Including the bulk vacuum energy $\Delta c_{\rm Bulk}$, we obtain
the dashed line. 
Finally correcting with the boundary vacuum energy $2\Delta c_{\rm boun}$
we can obtain the $c_{\rm eff}(R)$ graphs which are identical 
upto $R\sim{\cal O}(1)$ as shown in Fig.1.
%%%%%%%%%%%%%%%%%%%%%%%%

\subsection{$(+f)$ and $(f+)$ boundary conditions}

The fugacity for these two equivalent BCs is given by
\beq
\la^{(+f)}_{j}(\th)=R^{(+)}_j\left(b,\th+{i\pi\over{2}}\right)
R^{(+)}_j\left({1\over{b}},-\th+{i\pi\over{2}}\right)
\label{plusfree}
\eeq
and the reflection amplitudes by
$s_1=ib/2$ and $s_2=i/(2b)$.
These two BCs are equivalent since they are related by exchanging 
the left and right boundaries. 
Furthermore, they are self-dual as one can see from Eq.({plusfree}).
Therefore, we can consider $0<B<1/2$ only.
Using the same procedure, we can compare two results in Tables 4--6.
\begin{table}[t]
\centerline{
\begin{tabular}{||c||c||c|c|c||} \hline
\rule[-.4cm]{0cm}{1.cm}B
& $\de^{\rm (RA)}_1$ & $\de_1^{\rm (TBA)}(A_2)$
& $\de_1^{\rm (TBA)}(A_3)$ & $\de_1^{\rm (TBA)}(A_4)$ \\ \hline
0.20 & --8.40840 & --8.40846 & --8.40846 & --8.40822 \\
0.25 & --6.02901 & --6.02904 & --6.02904 & --6.02903 \\
0.30 & --4.47772 & --4.47773 & --4.47774 & --4.47773 \\
0.35 & --3.45078 & --3.45079 & --3.45079 & --3.45079 \\
0.40 & --2.79439 & --2.79440 & --2.79440 & --2.79439 \\
0.45 & --2.42719 & --2.42719 & --2.42719 & --2.42719 \\
0.50 & --2.30886 & --2.30886 & --2.30886 & --2.30886 \\ \hline
\end{tabular}
}
\caption{$\de_1^{\rm (RA)}$ vs.
$\de_1^{\rm (TBA)}$ for $A_2,A_3$, and $A_4$ ATFTs for $(+f)$ BC.}
\end{table}
\begin{table}
\centerline{
\begin{tabular}{||c||c||c|c|c||} \hline
\rule[-.4cm]{0cm}{1.cm}B
& $\de^{\rm (RA)}_3$ & $\de_3^{\rm (TBA)}(A_2)$
& $\de_3^{\rm (TBA)}(A_3)$ & $\de_3^{\rm (TBA)}(A_4)$ \\ \hline
0.20 & 47.6454 & 47.7694 & 47.7812 & 48.1325 \\
0.25 & 29.5225 & 29.5766 & 29.5815 & 29.5700 \\
0.30 & 19.1384 & 19.1616 & 19.1642 & 19.1608 \\
0.35 & 12.8599 & 12.8691 & 12.8707 & 12.8711 \\
0.40 & 9.08618 & 9.08921 & 9.09027 & 9.09219 \\
0.45 & 7.05398 & 7.05450 & 7.05537 & 7.05780 \\
0.50 & 6.41097 & 6.41084 & 6.41165 & 6.41421 \\ \hline
\end{tabular}
}
\caption{$\de_3^{\rm (RA)}$ vs.
$\de_3^{\rm (TBA)}$ for $A_2,A_3$, and $A_4$ ATFTs for $(+f)$ BC.}
\end{table}
\begin{table}
\centerline{
\begin{tabular}{||c||c||c|c|c||} \hline
\rule[-.4cm]{0cm}{1.cm}B
& $\de^{\rm (RA)}_5$ & $\de_5^{\rm (TBA)}(A_2)$
& $\de_5^{\rm (TBA)}(A_3)$ & $\de_5^{\rm (TBA)}(A_4)$ \\ \hline
0.20 & --419.960 & --422.929 & --422.818 & --420.097 \\
0.25 & --200.593 & --201.713 & --201.666 & --200.605 \\
0.30 & --102.893 & --103.327 & --103.316 & --102.898 \\
0.35 & --54.4153 & --54.5765 & --54.5799 & --54.4176 \\
0.40 & --29.2623 & --29.3138 & --29.3226 & --29.2635 \\
0.45 & --16.9759 & --16.9859 & --16.9966 & --16.9766 \\
0.50 & --13.2727 & --13.2721 & --13.2833 & --13.2733 \\ \hline
\end{tabular}
}
\caption{$\de_5^{\rm (RA)}$ vs.
$\de_5^{\rm (TBA)}$ for $A_2,A_3$, and $A_4$ ATFTs for $(+f)$ BC.}
\end{table}

In Fig.3, we also plot the scaling functions $c_{\rm eff}(R)$ as a
function of $R$ setting ${\overline m}=1$ by considering both
the bulk and boundary vacuum energies.
In particular, the boundary energy is given by
\beq
\Delta c_{\rm boun}=\Delta c_{\rm boun}(b)+\Delta c_{\rm boun}(1/b).
\eeq
This agreement is a nonperturbative proof of the duality conjecture.

\section{Concluding Remarks}
In this paper we have derived the reflection amplitudes of the
simply-laced ATFTs with integrable BCs by considering
the quantum mechanical reflections of the wave functional in the Weyl
chamber and show that the results are consistent with those from
the functional relation method \cite{Fateev}.
The quantization conditions arising from these amplitudes generate the
ground state energies which are compared with the boundary TBA
equations based on the bulk and boundary $S$-matrices.
The excellent agreements of the two different approaches 
provide the nonperturbative checks for
the conjectured boundary $S$-matrices of the simply-laced ATFTs with
$(+)$ BCs and its dual (free) BCs
where the conjectured boundary vacuum
energies play  the essential role for the agreement to the order $R^1$.

The two different approches based on the quantum mechanical reflections
and TBA  analysis should in principle provide a useful nonperturbative check
for different BCs such as all or some $\eta_i=-1$
whose boundary $S$-matrices are conjectured in \cite{DelGan}.
Boundary ATFTs with simply-laced Lie algebras other than $A$-series
can be also studied in this way to fix the ambiguity in the CDD factors. 
Another interesting problem is to relate the ``$L$-channel'' TBA 
which generates the boundary entropy to the boundary one-point
functions of the ATFTs.
We hope to publish these results in other publications.

\section*{\bf Acknowledgement}
We thank V. Fateev and Al. Zamolodchikov for valuable discussions
and Univ. Montpellier II, CNRS, and KIAS for hospitality.
The work of C.K. was supported by the BK21 project of the Ministry of
Education.  This work is supported in part by 
KOSEF 1999-2-112-001-5 (CA,CR) and MOST-99-N6-01-01-A-5 (CA).

\begin{figure}
\rotatebox{0}{\resizebox{!}{15cm}{\scalebox{0.1}{%
{\includegraphics[0cm,0cm][22cm,22cm]{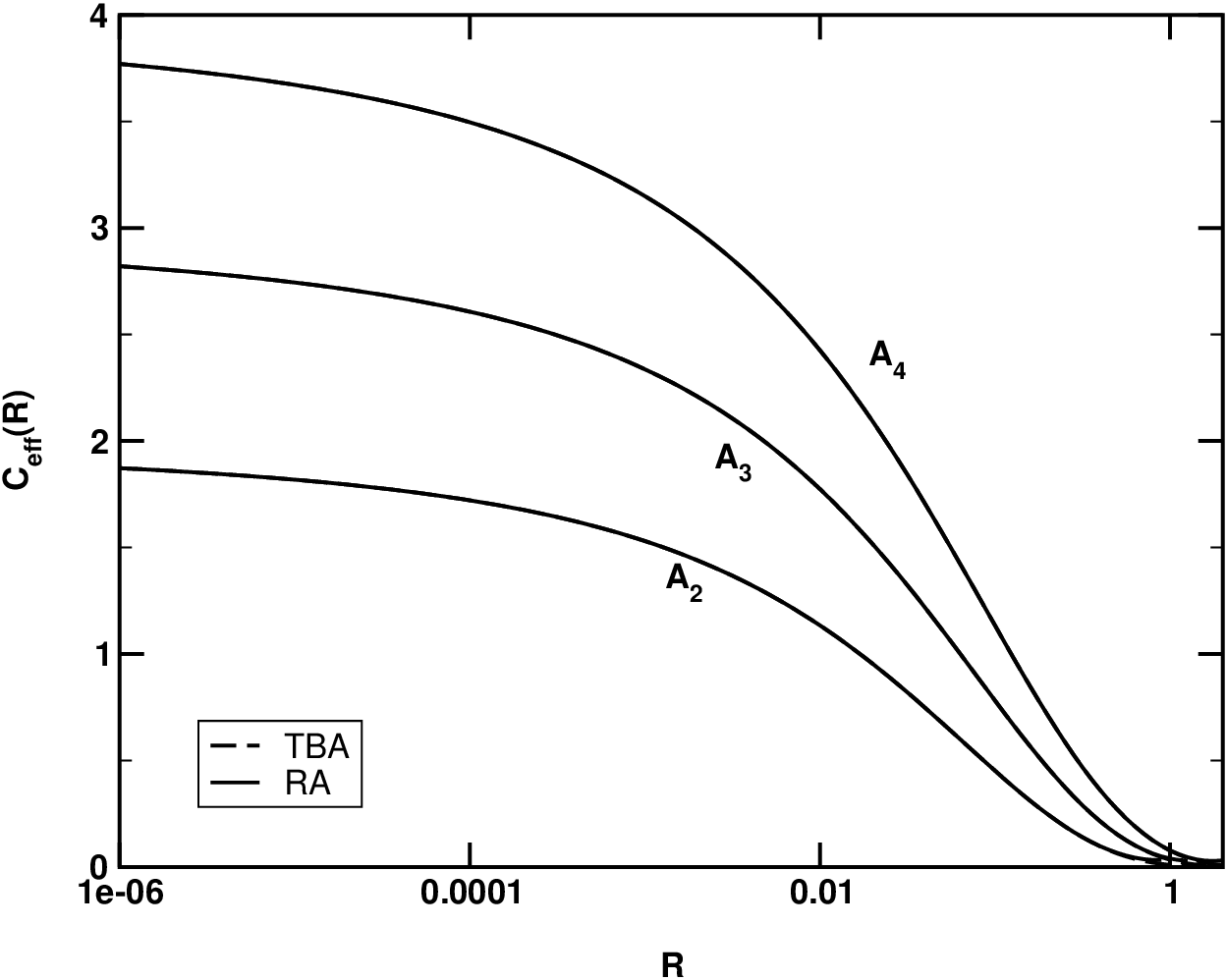}}}}}
\caption{Plot of $c_{\rm eff}$ for $A_2, A_3, A_4$ ATFTs at $B=0.4$
for $(++)$ BC.} 
\end{figure}

\begin{figure}
\rotatebox{0}{\resizebox{!}{15cm}{\scalebox{0.1}{%
{\includegraphics[0cm,0cm][22cm,22cm]{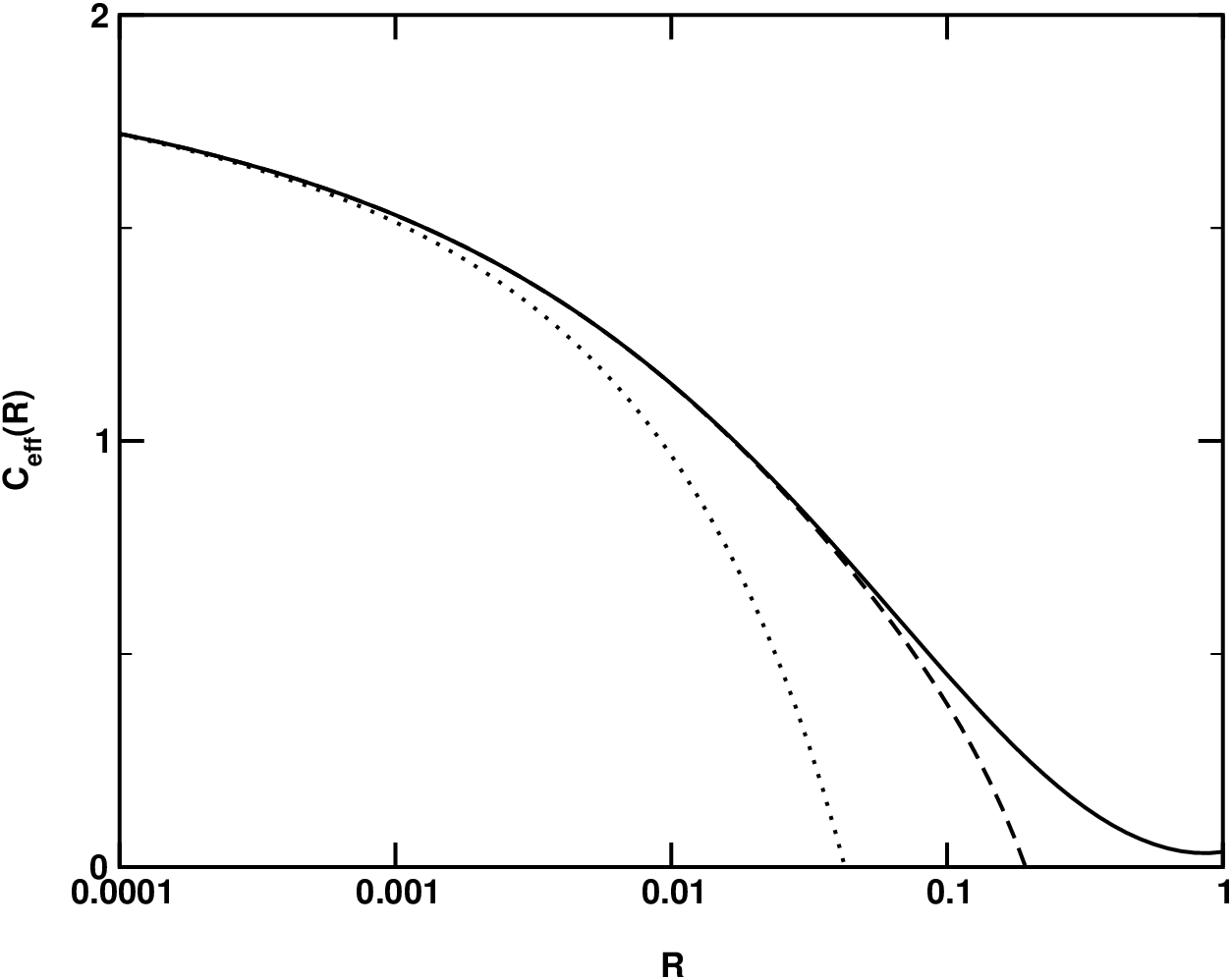}}}}}
\caption{Plot of $c_{\rm eff}$ for $A_2$ ATFTs for $(++)$ BC
without any vacuum energy (dotted line) and
with boundary vacuum energy (dashed line).
Including both energies (solid line), two results become identical
upto $R\sim{\cal O}(1)$.}
\end{figure}

\begin{figure}
\rotatebox{0}{\resizebox{!}{15cm}{\scalebox{0.1}{%
{\includegraphics[0cm,0cm][22cm,22cm]{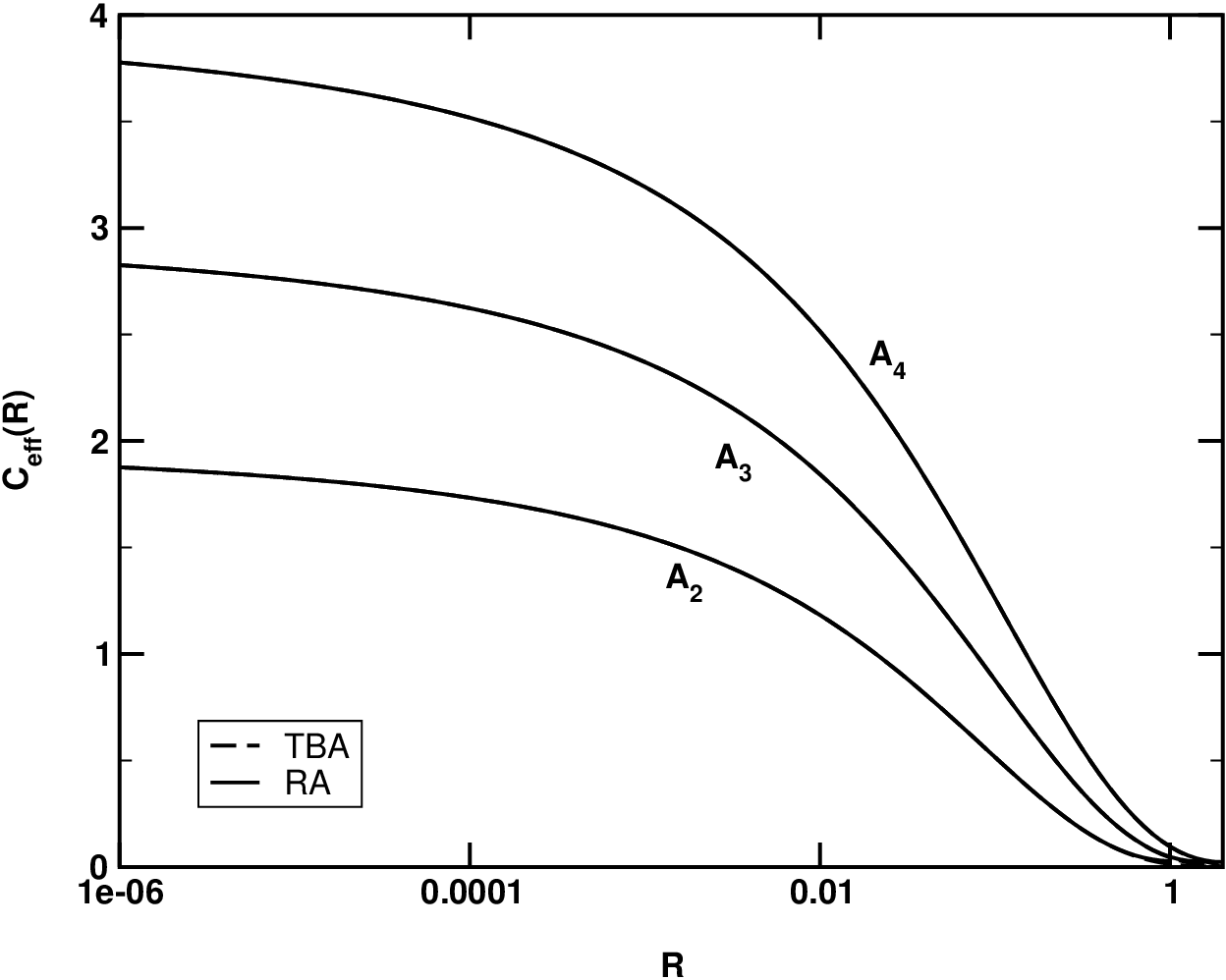}}}}}
\caption{Plot of $c_{\rm eff}$ for $A_2, A_3, A_4$ ATFTs at $B=0.4$
for $(+f)$ BC.}
\end{figure}

\end{document}